\documentclass[preprint]{aastex}

\shorttitle{The Redshift and Metallicity of the Host Galaxy of Dark GRB 080325 at z=1.78}
\shortauthors{Hashimoto et al.}
\begin{document}
\title{The Redshift and Metallicity of the Host Galaxy of Dark GRB 080325 at z=1.78}
\author{
Tetsuya Hashimoto\altaffilmark{1}, 
Daniel A. Perley\altaffilmark{2},
Kouji Ohta\altaffilmark{3}, 
Kentaro Aoki\altaffilmark{4}, 
Ichi Tanaka\altaffilmark{4}, 
Yuu Niino\altaffilmark{1},
Kiyoto Yabe\altaffilmark{1}, 
and Nobuyuki Kawai\altaffilmark{5}
}
\altaffiltext{1}{National Astronomical Observatory of Japan, 2-21-1 Osawa, Mitaka, Tokyo 181-8588, Japan}
\altaffiltext{2}{Department of Astronomy, California Institute of Technology, MC 249-17, 1200 East California Blvd, Pasadena, CA 91125, USA}
\altaffiltext{3}{Department of Astronomy, Kyoto University, Kyoto 606-8502, Japan.}
\altaffiltext{4}{Subaru Telescope, National Astronomical Observatory of Japan, 650 North A'ohoku Place, Hilo, HI 96720, USA.}
\altaffiltext{5}{Department of Physics, Tokyo Institute of Technology, 2-12-1 Ookayama, Meguro-ku, Tokyo 152-8551, Japan.}


\begin{abstract}
We present near-infrared spectroscopy of the host galaxy of dark GRB 080325 using Subaru/MOIRCS. 
The obtained spectrum provides a clear detection of H$\alpha$ emission and marginal [NII]$\lambda$6584. 
The host is a massive (M$_{*}\sim10^{11}$M$_{\odot}$), dusty ($A_{V}\sim 1.2$) star-forming galaxy at z=1.78. 
The star formation rate calculated from the H$\alpha$ luminosity (35.6-47.0 M$_{\odot}$ yr$^{-1}$) is typical among GRB host galaxies (and star-forming galaxies generally) at z $>$1; however, the specific star formation rate is lower than normal star-forming galaxies at redshift $\sim$ 1.6, in contrast to the high specific star formation rates measured for many of other GRB hosts. 
The metallicity of the host is estimated to be 12+log(O/H)$_{\rm KK04}$$=$8.88. 
We emphasize that this is one of the most massive distant host galaxies for which metallcity is measured with emission-line diagnostics.
The metallicity is fairly high among GRB hosts.
However, this is still lower than the metallicity of normal star-forming galaxies of the same mass at z$\sim$1.6.
The metallicity offset from normal star-forming galaxies is close to a typical value of other GRB hosts and indicates that GRB host galaxies are uniformly biased toward low metalicity over a wide range of redshift and stellar mass.
The low-metallicity nature of the GRB 080325 host is likely not attributable to the fundamental metallicity relation of star-forming galaxies beacuse it is a metal-poor outlier from the relation and has a low sSFR.
Thus we conclude that metallicity is important to the mechanism that produced this GRB.

\end{abstract}
\keywords{galaxies: spectroscopy --- gamma rays: burst}

\section{INTRODUCTION}
Long-duration Gamma Ray Bursts (hereafter, GRBs) are among most energetic phenomena in the Universe. 
The extremely luminous optical afterglows associated with these explosions can be bright enough to be visible even with the naked eye \citep[e.g.][]{2008Natur.455..183R} and are observable at very high redshifts \citep[$z>6$ and beyond, e.g.,][]{2006Natur.440..184K,2009Natur.461.1254T,2009Natur.461.1258S}. 
While it is now widely accepted that GRBs originate from death of massive stars, the identity of the progenitor and the evolutionary pathway required to produce it remain a matter of discussion.
Some theoretical models of this process involving the evolution of a single, rapidly-rotating massive star \citep[e.g.,][]{2006A&A...460..199Y,2006ApJ...637..914W} require a very low-metallicity environment, a possibility which can be tested by examining the properties of the host galaxy population. 
Indeed, many low-$z$ GRB host galaxies are low-luminosity and show blue colors \citep[e.g.,][]{2003A&A...400..499L}, suggesting that they are metal-poor, given the mass-metallicity relationship for star-forming galaxies \citep[e.g.,][]{2004ApJ...613..898T,2006ApJ...644..813E,2009ApJ...691..140H,2012PASJ...64...60Y,2014MNRAS.437.3647Y}. 
Spectroscopic observations of many of these GRB host galaxies confirm their low metallicities \citep[e.g.,][]{2006AcA....56..333S,2008AJ....135.1136M}. 

However, several examples of more massive, red GRB hosts have also been reported in recent years \citep[e.g.,][]{2007ApJ...660..504B,2010ApJ...719..378H,2011A&A...534A.108K,2011ApJ...736L..36H,2012MNRAS.421...25S,2013ApJ...778..128P}, most of which are associated with ``dark'' GRBs \citep[][]{2004ApJ...617L..21J}.  This hints at the possibility of a high-metallicity environment for at least some GRBs. 
In fact, high metallicities have been reported for a few cases based on emission-line diagnostics \citep[][]{2010ApJ...712L..26L,2012PASJ...64..115N,2013A&A...556A..23E,2013ApJ...774..119G,2014A&A...566A.102S}. 
These high-metallicity hosts seem to contradict models in which the GRB progenitor can \emph{only} form in metal-poor environments.   However, the average metallicity of a host galaxy does not necessarly reflect the metallicity of the GRB explosion site \citep{2011MNRAS.417..567N,2014arXiv1408.7059N}.  Metallicity measurements of both the host galaxy and the GRB explosion site are important to reveal a complete picture of GRB origins.

GRB 080325 is a dark GRB whose NIR afterglow and host galaxy were found through a target-of-opportunity program with the Subaru telescope \citep{2008GCN..7524....1T}. 
The host is a red, massive ($\sim$ 10$^{11}$M$_{\odot}$) galaxy at z$_{\rm phot}$ $\sim$ 1.9, as previously estimated by spectral energy distribution (SED) fitting analysis \citep{2010ApJ...719..378H}.
Thus, this GRB host serves as good test case to investigate the influence of physical environment in producing GRBs in the high redshift universe.

This paper is organized as follows. 
We present our near-infrared spectroscopy of the GRB 080325 host galaxy and spectral analysis in Section \ref{obs} and briefly describe results of our analysis in Section \ref{results}. 
In Section \ref{DISCUSSION}, we discuss properties of the GRB 080325 host galaxy, focusing in particular on the star formation rate (SFR) and metallicity and their variation between possible multiple components of the host. 
Finally our results and discussions are summarized in Section \ref{SUMMARY}.

Throughout this paper we use cosmological parameters of 
$H_{0}$ = 70.0 km s$^{-1}$ Mpc$^{-1}$, $\Omega_{M}$ = 0.279, and $\Omega_{\Lambda}$ = 0.721 \citep{2013ApJS..208...20B}.

\section{NEAR-INFRARED SPECTROSCOPY OF GRB 080325 HOST}
\label{obs}
\begin{figure}
\begin{center}
\includegraphics[scale=0.5]{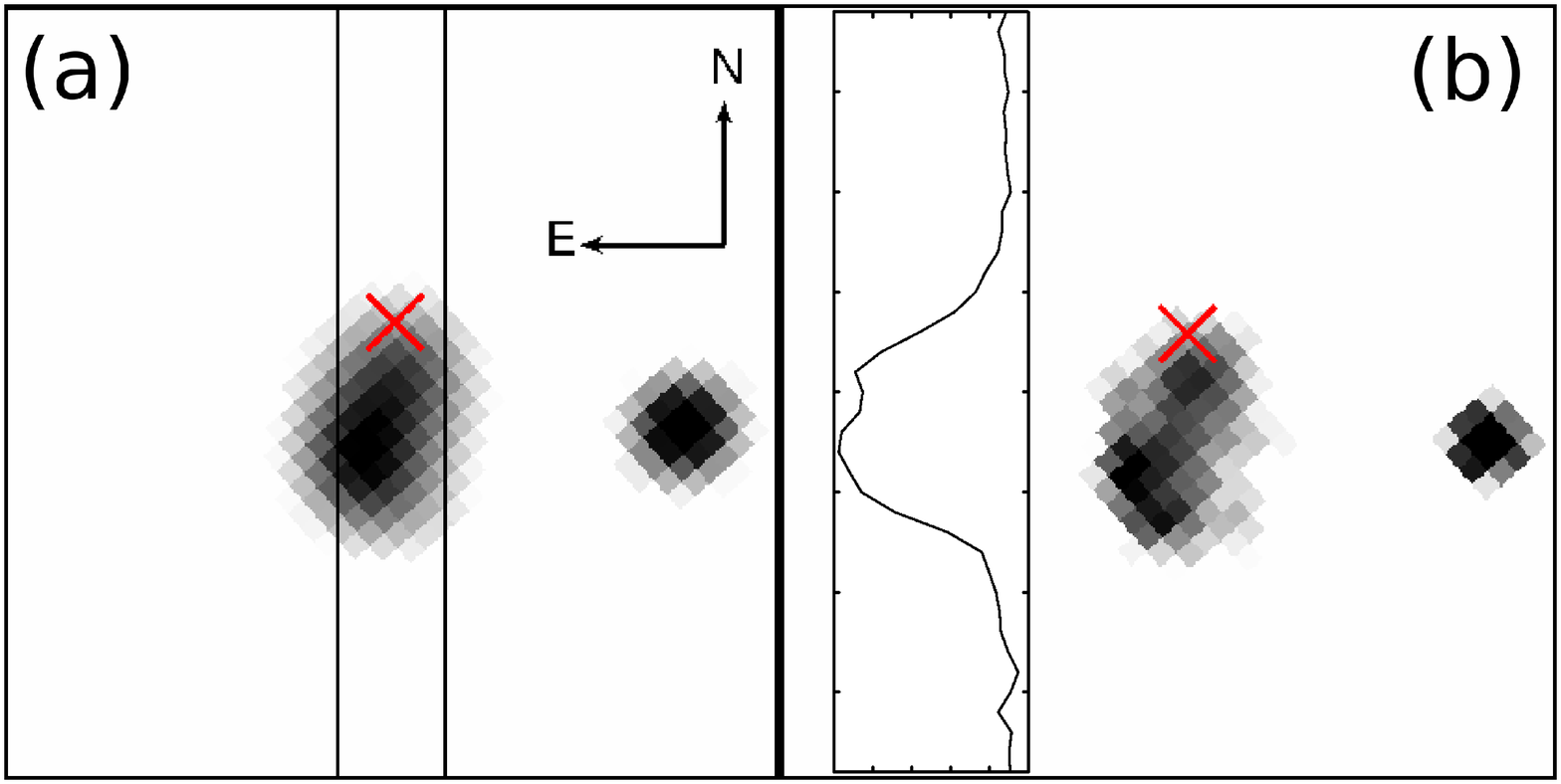}
\caption{
HST/WFC3 $J$-band images (5$\arcsec$.0 x 5$\arcsec$.0) of GRB 080325 host galaxy along with a neighboring point source. 
Image (a) is smoothed to have seeing size of $\sim 0\arcsec.6$ (the typical seeing during Subaru spectroscopic observations) for the purpose of the slit-loss estimation. 
The red cross represents the position of GRB afterglow \citep{2008GCN..7524....1T}.
The two black lines lines represent the position of the Subaru/MOIRCS 0$\arcsec$.7-slit.
Image (b) is the original HST/WFC3 $J$-band image with FWHM $\sim$ 0$\arcsec$.29 along with north-south profile of surface brightness of the host.
}
\label{HST_J} 
\end{center}
\end{figure}

We obtained a spectrum of the host galaxy of GRB 080325 using the
Subaru/Multi-Object Infrared Camera and Spectrograph \citep[MOIRCS;][]{2006SPIE.6269E..16I,2008PASJ...60.1347S} HK500 grism covering 1.3-2.5 $\mu$m as well as a standard star FS147 (A0) on 2011 June 22 and 23. 
Weather conditions were basically clear throughout the observation.
The seeing varied between approximately 0$\arcsec$.5-0$\arcsec$.8 in $K_s$-band. 
The slit (0$\arcsec$.7 width) was oriented to cover a large part of the host galaxy, as shown in Figure \ref{HST_J} (a); this width provides a resolving power of R = 630 at 1.8 $\mu$m. 
A standard ABBA sequence was employed to subtract OH sky emission lines 
with a total 7-hour exposure on source.

\begin{figure}
\begin{center}
\includegraphics[scale=0.6]{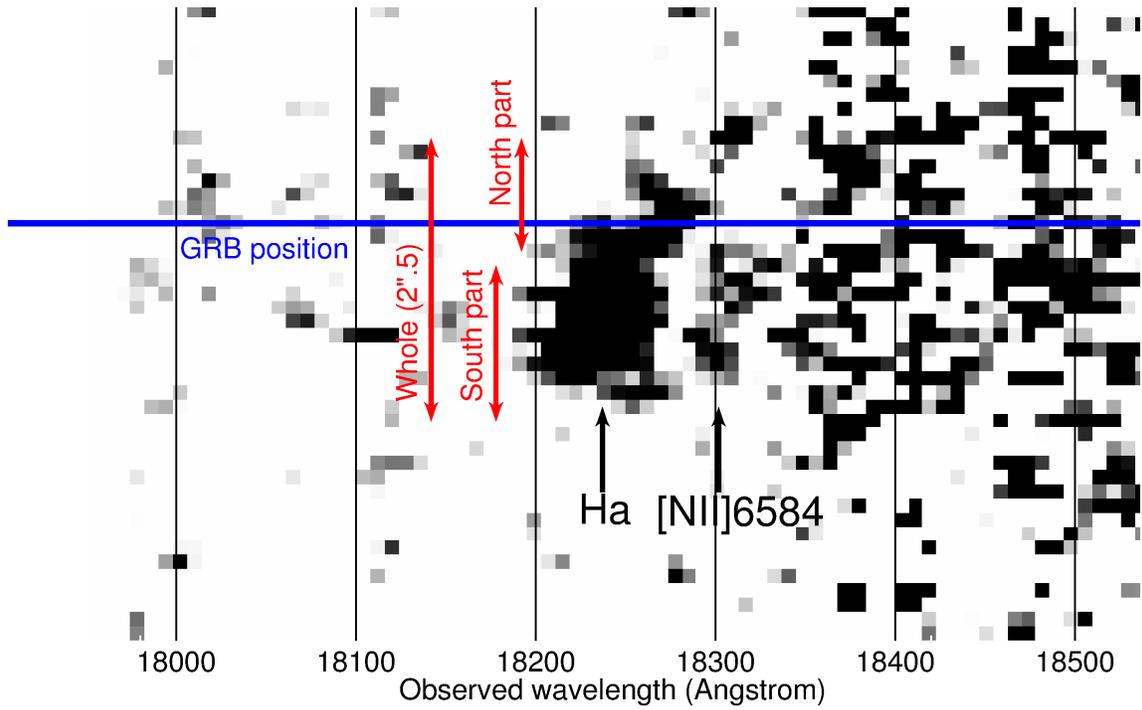}
\caption{
2D spectrum of GRB 080325 host galaxy obtained by MOIRCS/HK500 grism.  The blue horizontal line corresponds to the position of the afterglow reported by \cite{2008GCN..7524....1T}. Red arrows show the extracted regions used to make one-dimensional spectra.
}
\label{extract_Ha} 
\end{center}
\end{figure}

  We detected a strong, spatially- and spectrally-resolved emission line at a wavelength of $\lambda = 18243.4 \AA$, 
  as well as faint continuum at the expected position of the host galaxy (Figure \ref{extract_Ha}) on the trace.
  We associate this line with H$\alpha$ at $z=1.78$ on the basis of the absence of any other strong lines in our
  spectral range, a probable detection of [NII]$\lambda$ 6584 at the appropriate wavelength, and the good 
  consistency with the photometric redshift measurement (see Section \ref{TWO_COMPONENTS} for additional
  discussion).
The GRB position is denoted by a horizontal line in Figure \ref{extract_Ha}, which is calculated from the offset distance between the afterglow and the center of the host galaxy \citep{2010ApJ...719..378H}. 
The morphology of the H$\alpha$ line in the 2D spectrum shows a redshifted component around the GRB position. 
We divided the spectrum into ``South'' and ``North'' parts, bounding the position where the H$\alpha$ velocity largely changes around the GRB position. 
These two spectra roughly correspond to the two resolved components of the host galaxy seen in the $J$-band image from the Hubble Space Telescope (HST) obtained by \cite{2013ApJ...778..128P} as shown in Figure \ref{HST_J} (b).
We also simply summed these two spectra to extract the ``Whole'' spectrum of the host according to the conventional manner of many other previous GRB host studies.

\begin{figure}
\begin{center}
\includegraphics[scale=0.6, angle=-90]{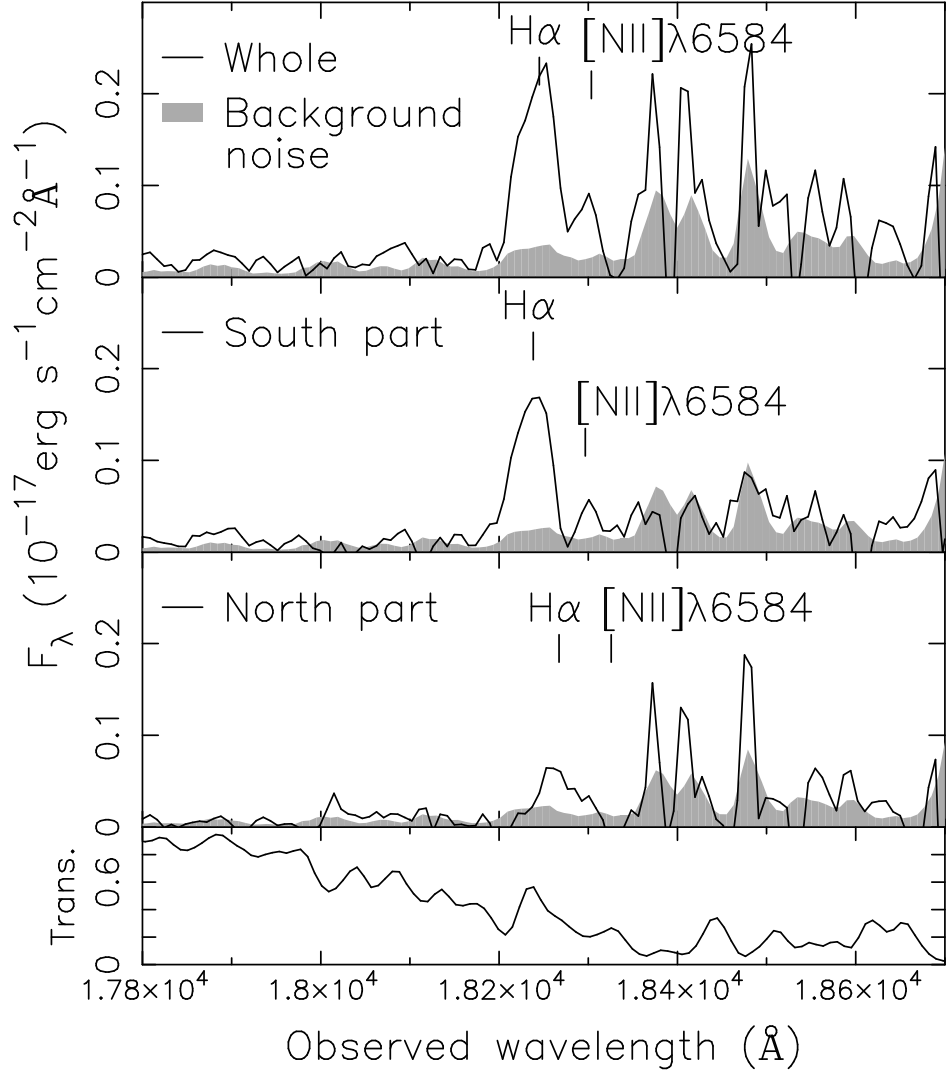}
\caption{
Spectra of GRB 080325 host galaxy extracted from Whole, North, and South parts as shown in Figure \ref{extract_Ha} and atmospheric transparency measured from a telluric standard star (bottom). 
Filled gray shows 1$\sigma$ background noise estimated from off-source region along the slit length. 
H$\alpha$ position and corresponding [NII]$\lambda$6584 are marked with vertical lines.
The pixel scale for the spectra is 7.93$\AA$/pix.
}
\label{spec} 
\end{center}
\end{figure}

Figure \ref{spec} shows spectra extracted from Whole, South, and North parts of the host galaxy along with the 1$\sigma$ background noise which is estimated from an off-source region along the slit length. 
The redshift of the GRB 080325 host is 1.780 as derived from Whole spectrum, which gives close agreement with the original photometric redshift within fitting uncertainties \citep{2010ApJ...719..378H}. 
The H$\alpha$ wavelength in the South spectrum corresponds to z=1.779.
Although the signal-to-noise ratio of H$\alpha$ in the North spectrum is poor, the redshift of North part is estimated to be 1.783. 
We also marginally detected the [NII]$\lambda$ 6584 line for Whole and South part spectra at about 3 $\sigma$ significance. 
We note that the positions of South and North parts of the host within the slit are slightly offset along a dispersion axis, although the two components are almost unresolved under actual observing condition as shown in Figure \ref{HST_J} (a). 
This may cause, to some extent, systematic difference of redshifts of each part of spectrum. 
The spatial offset roughly corresponds to $\delta \lambda$ = 11 $\AA$ on the detector, which is much less than the difference between the observed H$\alpha$ wavelengths in South and North parts (= 29 $\AA$).

\begin{figure}
\begin{center}
\includegraphics[scale=0.4, angle=-90]{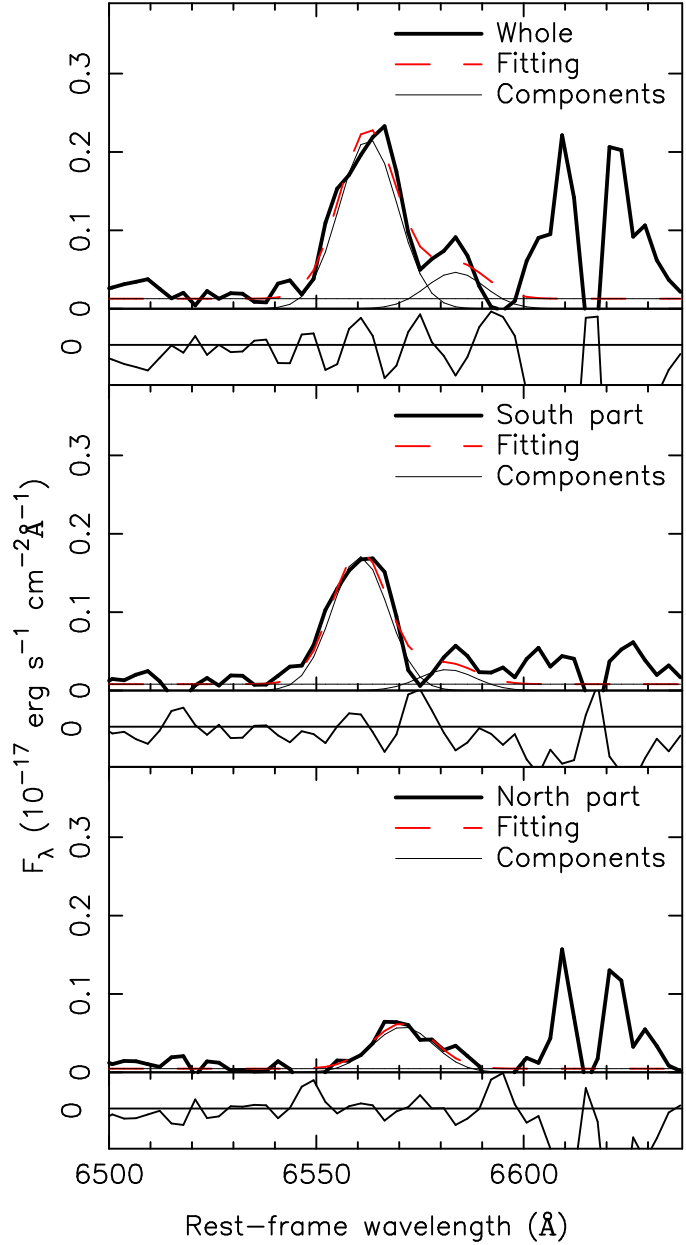}
\caption{
Spectral fits for spectra extracted from Whole, South, and North parts of the host galaxy of GRB 080325. 
The wavelength is converted to rest-frame given a redshift of z=1.78, as derived from H$\alpha$ emission in the Whole spectrum.
Red dashed lines are best-fit models and thin solid lines are each model components. 
Residuals between observed spectra and best-fit models are shown at the bottom of each spectrum.
}
\label{fit} 
\end{center}
\end{figure}

In order to measure emission-line fluxes, we performed spectral fitting analysis for each spectrum, assuming redshift and line width shared between H$\alpha$ and [NII]$\lambda$ 6584, their fluxes, and a constant continuum level as free parameters (Figure \ref{fit}). 
For North part, a single emission line and constant continuum are assumed. 
We used only the rest-frame wavelength range of 6400 - 6600$\AA$ for fitting analysis to avoid noisy background at $\lambda >$ 6600$\AA$.
Because the redshifted H$\alpha$ emission of North part spectrum could contaminate [NII]$\lambda$6584 flux of Whole spectrum, we also analysed the sum of South and North part spectra after shifting each spectrum to a common redshift based on the measured radial velocity offset of the H$\alpha$ line.
The kinematic and morphological multiple components of the host galaxy are discussed in Section \ref{TWO_COMPONENTS}.

\section{RESULTS}
\label{results}
We estimated metallicities for each spectrum using the [NII]$\lambda$6584/H$\alpha$ ratio based on \cite{2004MNRAS.348L..59P} method (hereafter PP04 N2). 
Since the absolute value of metallicity derived from emission-line diagnostics depends on the calibration method \citep[e.g.,][]{2003ApJ...591..801K,2008ApJ...681.1183K}, the metallicity comparison basically requires using an identical calibration. 
Here we adopt \cite{2004ApJ...617..240K} method (hereafter KK04) as a common metallicity calibration to compare with other metallicity measurements. 
The PP04 N2 metallicity of GRB 080325 host is converted to KK04 by using conversion formula parameterized by \cite{2008ApJ...681.1183K}. 
The metallicities of Whole, South, and North part spectra are 12+log(O/H)$_{\rm KK04}$ = 8.88, 8.78, and $<$8.75, respectively.

We also estimated the slit loss to be $\sim 0.3$ by smoothing the $J$-band HST image, i.e., $\sim 70 \%$ of the total flux from the host is incident within the $0\arcsec.7$ slit (Figure \ref{HST_J}).  
The total SFR of the host is 35.6-47.0 M$_{\odot}$ yr$^{-1}$ (see also Section \ref{STAR_FORMATION}), which is calculated from the H$\alpha$ luminosity using the conversion equation derived by \cite{2009ApJ...691..182S} and the slit loss. 

We also performed a full re-analysis of both Keck and Subaru optical photometry, using a common aperture radius of 1$\arcsec$.25 and a common field calibration from the Palomar 60-inch telescope for every image. 
The updated ground-based optical photometry magnitudes are summarized in Table \ref{mag}.
We performed a new SED fit with these optical data (fixing the redshift to the spectroscopic value of $z=1.78$) and NIR magnitudes reported in \cite{2010ApJ...719..378H} and \cite{2013ApJ...778..128P}, using a similar procedure as in \cite{2013ApJ...778..128P} but with a more flexible star-formation history model.
The best-fit result indicates that the host is a dusty ($A_{V}$ = 1.17 mag) massive (M$_{*}$ $\sim$ 10$^{11}$ M$_{\odot}$) star-forming galaxy, consistent with our previous estimates \citep[][]{2010ApJ...719..378H,2013ApJ...778..128P}. 
Metallicity, SFR, and some host properties derived from SED fitting analysis are summarized in table \ref{t_flux} along with results of sum of South and blue-shifted North part spectrum. 

\section{DISCUSSION}
\label{DISCUSSION}
\subsection{STAR FORMATION RATE}
\label{STAR_FORMATION}
\begin{figure}
\begin{center}
\includegraphics[scale=0.8,angle=-90]{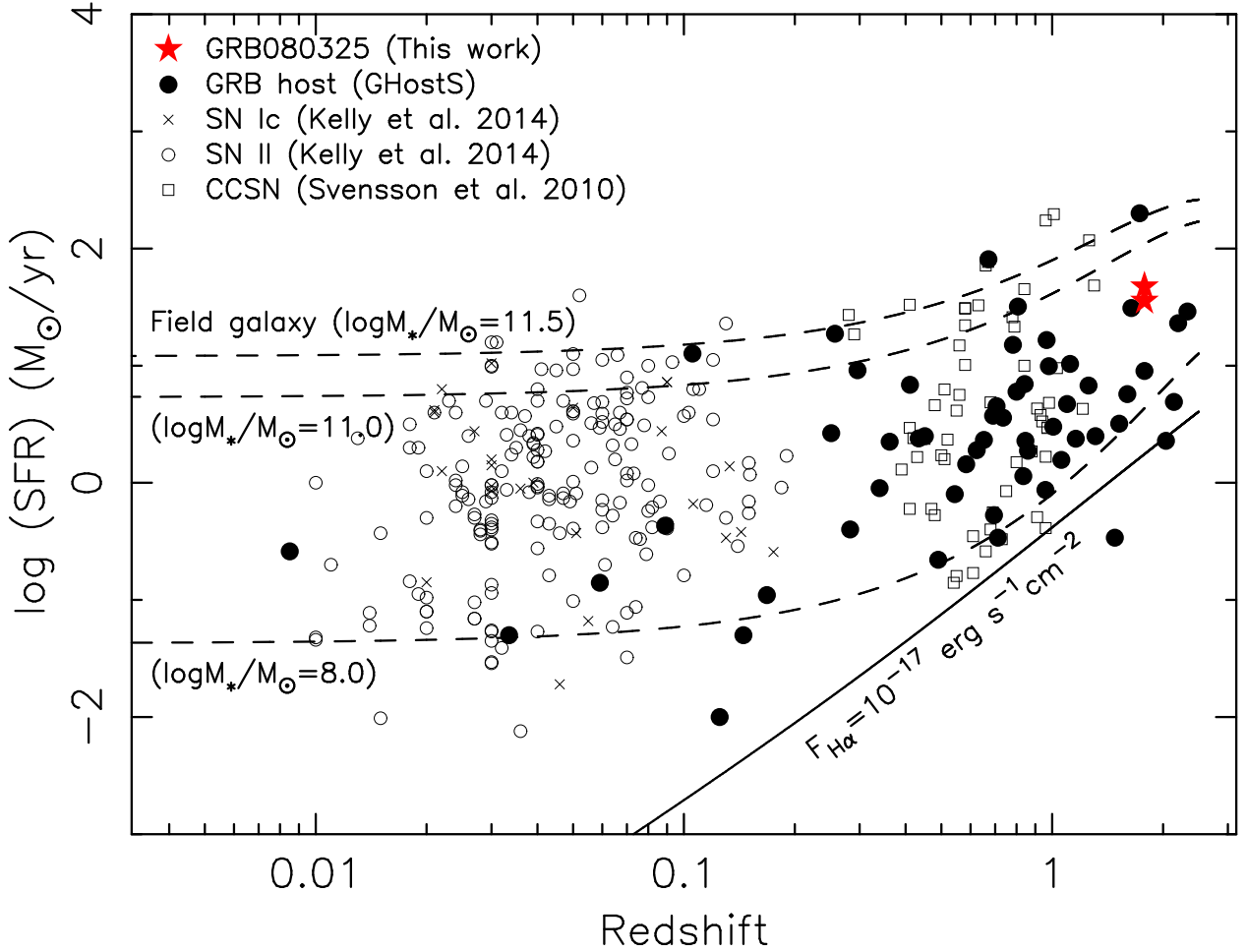}
\caption{
SFRs of GRB 080325 (red stars), assuming two cases of emission-line extinction of $A_{V,}$$_{\rm emission}$ = $A_{V,}$$_{\rm stellar}$ and $A_{V,}$$_{\rm emission}$ = $A_{V,}$$_{\rm stellar}$/0.76 \citep{2014ApJ...792...75Z} and other GRB hosts (black dots; GHostS project) as a function of redshift. 
For comparison, local core-collapse supernova hosts, i.e., type Ic (crosses) and II (open circle) hosts, are over plotted \citep{2014ApJ...789...23K} along with intermediate-redshift core-collapse supernova hosts \citep[squares;][]{2010MNRAS.405...57S}.
Three dashed lines correspond to the SFR of normal star-forming galaxies with stellar mass of log (M$_{*}$/M$_{\odot}$) = 8.0, 11.0, and 11.5, respectively (the star-formation ``main sequence''; \citealt{2012ApJ...754L..29W}). 
Solid line is SFR calculated from H$\alpha$ flux density, $f$ = 10$^{-17}$ erg s$^{-1}$ cm$^{-2}$ as a reference of detection limit of H$\alpha$.
}
\label{z_SFR} 
\end{center}
\end{figure}

The total SFR calculated from H$\alpha$ luminosity of the whole host galaxy (SFR$_{\rm H_{\alpha}}$) is 14.7 M$_{\odot}$ yr$^{-1}$ without any dust-extinction correction if a slit loss of $\sim 0.3$ (See Section \ref{obs}) is taken into account. 
The extinction-corrected SFR$_{\rm H_{\alpha}}$ is 35.6 M$_{\odot}$ yr$^{-1}$ assuming a \cite{2000ApJ...533..682C} extinction law, given the dust extinction ($A_{V}$ = 1.17 mag; table \ref{t_flux}) derived from SED fitting of the photometric data. 
This value is in agreement with the SFR estimated by the SED fitting (SFR$_{\rm SED}$ = 20.3$^{+13.1}_{-9.5}$).
However, the amount of extinction for the stellar component of a galaxy is not always same as for the emission lines \citep[e.g.,][]{2000ApJ...533..682C,2014ApJ...792...75Z}.
If such difference is taken into account (i.e., using an average correction factor of $A_{V,}$$_{\rm emission}$=$A_{V,}$$_{\rm stellar}$/0.76 at z $\sim$ 1.6; \citealt{2014ApJ...792...75Z}), SFR$_{\rm H_{\alpha}}$ is estimated to be 47.0 M$_{\odot}$ yr$^{-1}$. 
These SFRs are consistent with the distribution of other GRB host galaxies which show wide range, e.g., from $\sim$0.1 to $\sim$100 M$_{\odot}$/y at z $>$ 1 (Figure \ref{z_SFR}). 
In the figure, the GRB sample is collected from publicly available GHostS database\footnote{http://www.grbhosts.org}; core-collapse supernovae are taken from \cite{2010MNRAS.405...57S} and \cite{2014ApJ...789...23K}. 
Top and bottom dashed lines show the redshift evolution of SFR of filed star-forming galaxies with log(M$_{*}$/M$_{\odot}$)=8.0 and 11.5 \citep{2012ApJ...754L..29W}, although the uncertainty is large at higher redshift for less-massive galaxies. 
Because of the well-known correlation between SFR and stellar mass for star-forming galaxies (the so-called ``star formation main sequence''; \citealt[e.g.,][]{2004MNRAS.351.1151B,2007ApJ...660L..43N}), field star-forming galaxies distribute between two dashed lines.
Our result supports that the SFR distribution of GRB host galaxies falls in the same range as normal star-forming galaxies over a wide range of redshift up to $\sim$ 1.8.

\cite{2011MNRAS.414.1263M} demonstrated that the physical properties of GRB host galaxies are consistent with the fundamental metallicity relation (FMR) of normal star-forming galaxies---that is, the tight dependence of metallicity on stellar mass and SFR \citep{2010MNRAS.408.2115M}. 
They suggested that low metallicity is not necessarily important for galaxy to host a GRB but that GRB production seems to be related to efficiency of star formation as measured by the \emph{specific} SFR (sSFR), the SFR normalized to the total stellar mass of the galaxy.
In fact, previously researched low-z GRB host galaxies are biased toward higher sSFR compared with local star-forming galaxies \citep[e.g.,][]{2011MNRAS.414.1263M}.

For the GRB 080325 host, the SFR is \emph{lower} than for field star-forming galaxies with stellar mass $\sim$ 10$^{11}$M$_{\odot}$ at z $\sim$ 1.6 \citep[e.g.,][see also middle dashed line in Figure \ref{z_SFR}]{2013ApJ...777L...8K}. 
This means that the sSFR of the host galaxy is lower than normal for star-forming galaxies at a fixed stellar mass.
The sSFR of the host is (0.34-0.45) Gyr$^{-1}$ (table \ref{t_flux}), which is likewise lower than the typical value of $\sim$ 1.0 Gyr$^{-1}$ for normal star-forming galaxies with same stellar mass at z$\sim$ 1.6 \citep{2013ApJ...777L...8K}.
We also estimated the expected metallicity from the FMR derived by \cite{2010MNRAS.408.2115M} as follows.
\begin{equation}
\label{eq2}
12+{\rm log(O/H)_{FMR}}=8.90+0.39x-0.20x^{2}-0.077x^{3}+0.064x^{4},
\end{equation}
where $x=\mu_{0.32}-10.0$, $\mu_{0.32}={\rm log}(M_{*}/M_{\odot})-0.32 \times {\rm log}({\rm SFR})$. 
The expected metallicity is $12+{\rm log(O/H)}_{\rm FMR}$ $\sim$ 9.0. 
Even if the systematic metallicity calibration error ($\sim$ 0.1) is taken into account, this is higher than our spectroscopically measured value of $12+{\rm log(O/H)}_{\rm N06}$ = 8.7, as estimated from the [NII]$\lambda$6584/H$\alpha$ ratio. 
Note that the \cite{2006A&A...459...85N} calibration (N06) is used here, in common with the calibration for $12+{\rm log(O/H)}_{\rm FMR}$.

The GRB 080325 host therefore has relatively low sSFR and is an outlier from the FMR, in contrast with most low-z GRB host galaxies. 
Although the redshift dependency of FMR is under debate \citep[][]{2010MNRAS.408.2115M,2012PASJ...64...60Y,2014MNRAS.437.3647Y,2014ApJ...792...75Z}, this suggests that the presumption of \cite{2011MNRAS.414.1263M}, i.e., high star-forming efficiency plays an important role in production of GRB rather than low metallicity, is not always applicable to high-z GRB hosts. 

\subsection{METALLICITY OF THE GRB 080325 HOST}
\label{METALLICITY}
The host galaxy of GRB 080325 is one of the most massive and most distant host galaxies for which a metallicity has been determined by emission-line diagnostics (See Section \ref{results} and table \ref{t_flux} for the detailed description of metallicity measurement). 
This metallicity should be compared with other galaxies at similar, high redshifts when the cosmic star formation rate density was at its peak, in contrast to previous GRB host emission-line studies, nearly all of which are at z $<1.0$ (except for GRB 080605 at z=1.64; \citealt{2012A&A...546A...8K} and GRB 121024A at z=2.3; \citealt{2014arXiv1409.6315F}). 
Figure \ref{mass-metal} shows the mass-metallicity relation for a sample of GRB host galaxies; observations are mainly collected by \cite{2014PASP..126....1L} and supplemented with stellar masses derived by \cite{2010MNRAS.405...57S} and \cite{2014A&A...566A.102S}. 
We added to this 
GRB 011121 \citep[][]{2010MNRAS.405...57S,2013ApJ...774..119G}, 
GRB 060505 \citep[][]{2008ApJ...676.1151T,2013ApJ...774..119G}, 
GRB 080605 \citep{2012A&A...546A...8K}, 
GRB 100418A \citep[][]{2012PASJ...64..115N}, 
GRB 110918A \citep{2013A&A...556A..23E}, 
GRB 120422A \citep{2014A&A...566A.102S}, 
and 121024A \citep{2014arXiv1409.6315F} 
as well as the GRB 080325 host galaxy (this paper). 
Although the possibility of a systematic offset between metallicities derived from emission-line diagnostics and absorption-line systems of GRB afterglow is still unclear, absorption-line metallicities at z $<$ 4 are also shown in the figure just for reference (metallicity measurements from \citealt{2014arXiv1408.3578C} and stellar masses from GHostS and \citealt{2011ApJ...739....1L}.) 
For comparison, nearby type Ic and II supernova host galaxies \citep{2012ApJ...759..107K,2013ApJ...774..119G,2014ApJ...789...23K} are also denoted as well as mass-metallicity relations of field star-forming galaxies at various redshifts \citep{2014ApJ...791..130Z}. 
All stellar masses in the figure are converted as necessary using a \cite{2003PASP..115..763C} initial mass function and all metallicities are based on the KK04 method.
GRB 080325 host has a fairly high metallicity environment among GRB samples. 
This result suggests a disagreement with the presence of critical cutoff above which GRBs cannot occur (as proposed by e.g.\citealt{2008AJ....135.1136M}), even in the high-redshift universe. 
The metal-rich host galaxy also seems not to match up with the low-metallicity requirement suggested from theoretical modelling of single massive stellar evolution \citep[e.g.,][]{2006A&A...460..199Y,2006ApJ...637..914W}. 
One possibility to explain high-metal GRB hosts is an alternate progenitor scenario, such as a binary system \citep{1995PhR...256..173N,1999ApJ...526..152F,2000ApJ...534..660I} or magnetar. 
As another possibility, we note that metallicity measurement does not always reflect the immediate environment of GRB (even in spatially-resolved spectra such as ours) due to spectroscopic dilution by limited spatial resolution and poor signal-to-noise ratio unless at least spatial resolution of $\sim$ 500 pc is achieved \citep{2014arXiv1408.7059N}. 
Such requirement is not easy for high-redshift GRBs, even if GRB-site spectrum is extracted from the host. 
There is a hint that the metallicity upper limit for GRB 080325 site is lower than metallicity of whole host but the spatial resolution is not high enough to examine the immediate environment of GRB. 
The HST image suggests that the host system may be a major merger (See Section \ref{TWO_COMPONENTS} for the detailed discussion), in which the GRB is associated with northern stellar component.
Because of this situation, the discovery of a high-metal host does not rule out locally low-metal environment around the GRB. 

The GRB 080325 host is metal-rich but its metallicity is still lower than field star-forming galaxies at similar redshift (i.e., COSMOS sample denoted by a red solid line). 
Redshift evolution of the mass-metallicity relation \citep[e.g.,][]{2005ApJ...635..260S, 2012PASJ...64...60Y, 2014MNRAS.437.3647Y, 2014ApJ...791..130Z} and the wide redshift distribution of the GRB sample often make this kind of comparison confusing. 
To clarify this, we adopt the stellar mass normalized by $M_{o}$, the turnover mass of the mass-metallicity relation proposed by \cite{2014ApJ...791..130Z}, above which the metalicity asymptotically approaches the upper limit. 
Here log(M$_{o}$/M$_{\odot}$) is 9.12, 9.52, 9.81, and 10.11 for SDSS \citep[z=0.08;][]{2009ApJS..182..543A}, SHELS \citep[z=0.29;][]{2005ApJ...635L.125G}, DEEP2 \citep[z=0.78;][]{2003SPIE.4834..161D}, and COSMOS \citep[z=1.55;][]{2014ApJ...792...75Z} surveys respectively. 
They clearly demonstrated the mass-metallicity relation at z $<$ 1.6 is fairly independent of redshift if normalized galaxy stellar mass is used. 
They also showed that $M_{o}$ as a function of redshift is well fitted with a linear function as follows.

\begin{equation}
{\rm log}(M_{o}/M_{\odot})=(9.138\pm0.003)+(2.64\pm0.05){\rm log}(1+z)
\label{eq1}
\end{equation}

The stellar mass of Figure \ref{mass-metal} is converted to mass in the $M_{o}$ unit by using equation (\ref{eq1}), slightly extrapolating this relation to z=1.78. 
The normalized mass-metallicity relation is shown in Figure \ref{Tmass-metal}. 
The relations for the SDSS, SHELS, DEEP2, and COSMOS samples are drawn, but they are almost degenerate in the figure because the redshift evolution between the samples has largely been removed by the $M_{o}$ correction. 
The distribution of core-collapse supernova hosts is consistent with the distribution of normal field star-forming galaxies. 
On the other hand, GRB host galaxies including GRB 080325 host are clearly below the normalized mass-metallicity relation.
The offset metallicity (which we define as the difference between the observed metallicity and the metallicity expected given the host stellar mass given the normalized mass-metallicity relation) is shown in the bottom of the figure.  
The offset of the GRB 080325 host is rather close to the typical value of the GRB sample.
The distribution of GRB hosts is $\sim$ 0.4 dex lower than field galaxies without any clear dependency on normalized stellar mass. 
This indicates GRBs preferably occur in low-metal galaxies, uniformly biased toward typically $\sim$ -0.4 dex regardless of both redshift and host-galaxy stellar mass.
This tendency is also demonstrated in Figure \ref{z-offset}, which shows the metallicity offset as a function of redshift and stellar mass. 
While the direct comparison with emission-line and absorption-line metallicities probably include large uncertainty, offsets for GRB absorption-line metallicity measurements at z $<$ 4 are also shown in the figure for reference (that is, the offset from the mass-metallicity relation for star-forming galaxies at z=2.3-3.5 from \citealt{2008A&A...488..463M}.) 
There is no significant correlation between metallicity offset and redshift or between metallicity offset and stellar mass for GRB host galaxies, even if absorption-line metallicities based on GRB afterglows are included. 
This result supports the conclusions of by \cite{2010AJ....140.1557L}, who reported a mass-metallicity relation for GRB host galaxies shifted toward low metallicity by -0.42 dex (where they simply divided their sample into two redshift categories, i.e., z$<$0.3 and 0.3$<$z$<$1.0). 
We also emphasize that GRB 080325 is an important case demonstrating that a GRB can occur in low-metal galaxy compared with normal star-forming galaxies of the same mass in a similar manner to local GRBs, even in the high-redshift (beyond z$\sim$ 1) universe. 
We conclude that the low metallicity nature of the GRB 080325 host is probably not attributable to the FMR of star-forming galaxies because the host is an outlier of the relationship as discussed in Section \ref{STAR_FORMATION}; rather, the metallicity itself is probably essential for production mechanism of this GRB.
Recently \cite{2014arXiv1407.4456P} found that approximately 15 $\%$ of all GRBs occur in the most luminous galaxies such as submillimeter galaxies and ultra-luminous infrared galaxies and that these GRB hosts have stellar masses significantly lower than IR/submillimeter-selected field galaxies of similar luminosities.
They suggest that GRB rate may be suppressed in metal-rich environments but independently enhanced in high efficient star-formation. 
If this is broadly true for other GRB host galaxies, GRB 080325 host might be a distinct case, i.e., only the low metallicity is important to the production mechanism regardless of the sSFR.

According to spectroscopic surveys of high-redshift star-forming galaxies, there are indications that distant star-forming galaxies occupy a region of the BPT plane \citep{1981PASP...93....5B} distinct from star-forming galaxies in the local universe \citep[e.g.,][]{2005ApJ...635.1006S,2006ApJ...644..813E,2012PASJ...64...60Y,2014MNRAS.437.3647Y,2014ApJ...795..165S}. 
Recently, results from an extensive spectroscopic survey of high-redshift galaxies at 2.0 $<$ z $<$ 2.6 have been reported by \cite{2014ApJ...795..165S}.
They suggest that high-redshift galaxies have harder stellar ionizing radiation, higher ionization parameter, and shallower dependence of N/O on O/H than is typically inferred for galaxies in the local universe. 
Therefore it is not clear whether the metallicity calibrations based on HII regions in the local universe can be applied to high redshift universe in general.
It is actually found that oxygen abundance of PP04 N2 method is systematically higher than that of PP04 O3N2 method for high redshift galaxies \citep{2014MNRAS.437.3647Y,2014ApJ...795..165S}, which could be result from systematically higher N/O at a given O/H in the high redshift sample as discussed by \cite{2014ApJ...795..165S}. 
We note that even if N2 method gives systematically higher oxygen abundances in the high-redshift universe, our observations of the GRB 080325 host are still indicative of a low metallicity. 

While many of our conclusions are based on analysis of a marginally detection of the [NII]$\lambda$6584 emission line, we emphasize the conclusions all still hold even if [NII]$\lambda$6584 detection of GRB 080325 host is treated as an upper limit. 
The host still shows moderate SFR, low sSFR, and low metallicity compared with normal star-forming galaxies with same mass at z$\sim$ 1.6.

\subsection{TWO COMPONENTS OF HOST GALAXY?}
\label{TWO_COMPONENTS}
As shown in Figure \ref{extract_Ha}, H$\alpha$ emission from the North component of the host galaxy is redshifted by 474.5 km/s relative to the South part. 
This velocity offset is rather higher than a typical galactic rotation velocity at z $\sim$ 2 \citep[e.g.,][]{2009ApJ...706.1364F}, suggesting a merging system of multiple galaxies. 
In fact the $J$-band image (rest-frame $\sim$ 4400 $\AA$; Figure \ref{HST_J}) obtained by HST shows a double-peaked stellar component within the host. 
The position of the north stellar component in the HST image roughly corresponds to the North part spectrum which we defined kinematically. 
If this GRB occurred in the merging system of two galaxies, GRB is likely hosted by the north stellar component (although the possibility that it occurred in the northern edge of the south stellar component can not be excluded). 
Considering this situation, the stellar mass of the ``true'' host could be less massive than the total stellar mass of the combined system, which would make the discussion in Sections \ref{STAR_FORMATION} and \ref{METALLICITY} more complicated. 
In order to roughly estimate the individual stellar mass of the North component we divided the HST $J$-band image of the host into two components and measured a light fraction of North component. 
This fraction is $\sim 40 \%$ of the total light in the J band
, which suggests a comparable fraction of the total stellar mass is also in this component.
Given this mass for North stellar component and our earlier measurement of the SFR for the North part spectrum, the sSFR inferred for the North component is 0.32 Gyr$^{-1}$, which is still lower than typical star-forming galaxies at similar redshift \citep[e.g.,][]{2012ApJ...754L..29W}.
In a similar way, the upper limit on the metallicity derived from North part spectrum is 12+log(O/H) = 8.52 (N06 calibration for comparison with FMR metallicity) or 8.74 (KK04 calibration in Figure \ref{mass-metal}). 
This is also lower than the metallicity inferred from FMR ($\sim$ 9.0) or normalized mass-metallicity relation (= 8.98 in Figure \ref{mass-metal}) of normal star-forming galaxies, calculated by using North-part SFR and stellar mass.

Finally we comment on the possibility of misidentification of the emission lines.
We first note that the photometric redshift of 1.8 $<$ z$_{\rm phot}$ $<$ 2.2 \citep{2010ApJ...719..378H} rules out the association of our putative H$\alpha$ line with any other strong nebular emission lines at different redshifts: specifically, the cases of [OIII]$\lambda$5007 of a galaxy at z=2.65, [OIII]$\lambda$4959 at z=2.68, H$\beta$ at z=2.76 and [OII]$\lambda$3727 at z=3.90 are all strongly excluded by the photometric fit; the strong $U$ and $B$ detections in particular give no suggestion of a break indicating the onset of the Ly$\alpha$-forest as expected for a higher-redshift galaxy.  The marginal detection of [NII]$\lambda$6584 at the correct wavelength further supports our preferred redshift.  It is possible, although statistically very unlikely, that the North component may represent a background galaxy at one of these alternate redshifts behind an unrelated object at z=1.78; our photometric analysis would not be able to rule this out since the two objects are strongly blended except in the HST imaging.  However, even in this case we would expect to detect additional lines elsewhere in our spectral coverage; in particular, if it were to be one member of [OIII]$\lambda$5007, [OIII]$\lambda$4959, or H$\beta$ then we would expect to observe the other lines (as well as [OII]$\lambda$3727) and in all cases do not.   In the case of [OII]$\lambda$3727 at z=3.90, all other expected strong nebular emission lines are out of range, but the chance of  a sub-arcsecond positional association of two galaxies with a redshift offset precisely tuned to align two unrelated emission lines in observed wavelength nevertheless is extremely remote.  We therefore conclude that 
the emission line detected in the North spectrum is likely H$\alpha$ from a galaxy component physically associated with South component of the host. 

Thus we conclude that the kinematic and morphological complexity of the host does not significantly affect discussions on the low-metallicity nature of GRB hosts in Sections \ref{STAR_FORMATION} and \ref{METALLICITY}. 

\section{SUMMARY}
\label{SUMMARY}


We detected H$\alpha$ and [NII]$\lambda$6584 emission lines of the host galaxy of dark GRB 080325 by using Subaru/MOIRCS.
The host is a massive ($\sim$ 10$^{11}$M$_{\odot}$) dusty ($A_{V}$=1.17 mag) star-forming galaxy at z=1.78, in contrast to blue less massive GRB host galaxies at local universe.

The SFR indicated from H$\alpha$ is between 35.6-47.0 M$_{\odot}$ yr$^{-1}$ (depending on the ratio of emission-line extinction to stellar extinction), consistent with the SFR derived from SED fitting of the photometry. 
This value is typical among GRB host galaxies and supports that the SFRs of GRB host galaxies are distributed similarly to those of normal star-forming galaxies over a wide range of redshift up to at least z $\sim$ 1.8. 
On the other hand, the sSFR of the host is lower than normal for star-forming galaxies, in contrast with the high sSFR nature observed for many other GRB host galaxies.
In addition, the expected host metallicity calculated from the FMR of normal star-forming galaxies is higher than the actual metallicity measurement by using emission-line diagnostics, even if systematic error of metallicity calibration is taken into account. 
The GRB 080325 host is therefore both an outlier from the FMR (with a lower metallicity than expected given its mass and SFR) and has a low sSFR.   Although the evolution of FMR is still under discussion, this result suggests that the previous presumption of importance of high star-forming efficiency rather than low metallicity as the primary condition needed to produce a GRB is out of accordance with this case. 

The host metallicity derived from [NII]$\lambda$6584/H$\alpha$ is fairly high compared to GRB host galaxies at lower redshift, providing evidence against the existence of a critical metallicity cutoff above which GRBs never occur. 
Even if the cutoff existed, it is much higher than that suggested for low-z GRB host galaxies.
This may favor progenitor scenarios other than the canonical single-star model, such as models involving a binary system or magnetar. 
Another possibility is spectroscopic dilution due to limited spatial resolution.
Actually there is a hint of a local low-metal environment, i.e., the upper limit of metallicity at the GRB-site is lower than that of whole host galaxy, although the spatial resolution is not enough to investigate the specific HII region in which the GRB occurred.
 
In any case, the metallicity of the host is still lower than normal for massive star-forming galaxies at z$\sim$ 1.6.
To avoid confusion due to the wide redshift range of the current GRB sample and redshift evolution of mass-metallicity relation for normal star-forming galaxies, we adopted a normalized mass-metallicity relation that is independent of redshift. 
The metallicity offset of the host from the normalized mass-metallicity relation is $\lesssim$ -0.2 dex, which is close to typical value of other GRB hosts. 
We also found that the metallicity-offset distribution for GRB hosts is uniformly biased toward low metallicity regardless of redshift (0 $<$ z $<$ 1.8) and stellar mass (10$^{8}$ $<$ M$_{*}$/M$_{\odot}$ $<$ 10$^{11}$), compared with core-collapse supernova host galaxies.
GRB 080325 is an important case of a GRB occuring in a galaxy that is metal-poor compared with normal star-forming galaxies, even though it is a massive galaxy and even though it is at relatively high-redshift. 
We emphasize that the low-metallicity nature of GRB 080325 is likely not attributable to the FMR of star-forming galaxies since this is an outlier of FMR, i.e., low metallicity (not high sSFR) is likely essential for this burst.


\acknowledgments
We would like to thank the Subaru Telescope staff for their invaluable help for our observation with Subaru/MOIRCS. 
Work by T. Hashimoto  was supported by funding from National Astronomical Observatory of Japan and TMT project office in Japan. 
T. Hashimoto would like to acknowledge Department of Astronomy, California Institute of Technology to provide an excellent academic research environment for writing the paper.
T. Hashimoto also thanks to S. R. Kulkarni and T. Usuda to create an opportunity of this collaborative study.

\clearpage

\begin{figure}
\begin{center}
\includegraphics[scale=0.8, angle=-90]{mass-metal.eps}
\caption{
Mass-metallicity relation of GRB host galaxies. 
The red star is the metallicity of the GRB 080325 host galaxy (this work).
Beginning at the top, the 3 stars correspond to metallicies estimated from Whole, South, and North part spectra, respectively.
Black dots are GRB host samples mainly collected by \cite{2014PASP..126....1L}, supplementing with stellar mass derived by \cite{2010MNRAS.405...57S} and \cite{2014A&A...566A.102S} with additional samples of 
GRB 011121 \citep[][]{2010MNRAS.405...57S,2013ApJ...774..119G},
GRB 060505 \citep[][]{2008ApJ...676.1151T,2013ApJ...774..119G}, 
GRB 080605 \citep{2012A&A...546A...8K}, 
GRB 100418A \citep{2012PASJ...64..115N}, 
GRB 110918A \citep{2013A&A...556A..23E}, 
GRB 120422A \citep{2014A&A...566A.102S},
and 121024A \citep{2014arXiv1409.6315F}.
Filled squares are absorption-line metallicities of GRB afterglows at z $<$ 4 collected by \cite{2014arXiv1408.3578C} with stellar mass from GHostS and \cite{2011ApJ...739....1L}.
Each curved line shows the mass-metallicity relation for field star-forming galaxies at each redshift \citep{2014ApJ...791..130Z}. 
Error bars include fitting errors and the typical systematic error of metallicity calibration $\sim 0.1$ dex \citep{2008ApJ...681.1183K}. 
}
\label{mass-metal}
\end{center}
\end{figure}

\begin{figure}
\begin{center}
\includegraphics[scale=0.5, angle=-90]{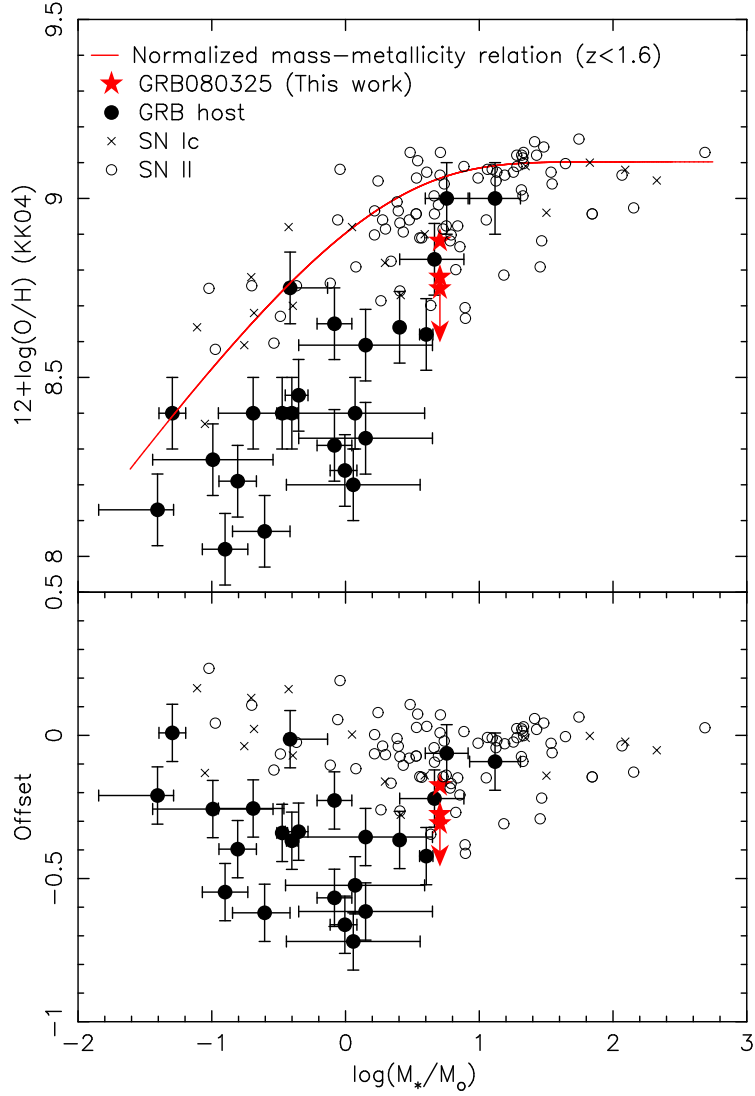}
\caption{
Same as figure \ref{mass-metal} but galaxy stellar mass is normalized by $M_{o}$ and displayed sample is limited to z $<$ 2.0 because $M_{o}$ is calculated for z$\lesssim$ 2 sample by \citealt{2014ApJ...791..130Z} (top). 
$M_{o}$ is the characteristic turnover mass of the mass-metallicity relation, above which the metallicity asymptotically approaches the upper limit. 
The red solid line is the normalized-mass metallciity relation for normal star-forming galaxies at z $<$ 1.6 \citep{2014ApJ...791..130Z}.
The offset metallicity of the GRB sample is defined relative to the normalized mass-metallicity relation (bottom). 
}
\label{Tmass-metal}
\end{center}
\end{figure}

\begin{figure}
\begin{center}
\includegraphics[scale=0.8, angle=-90]{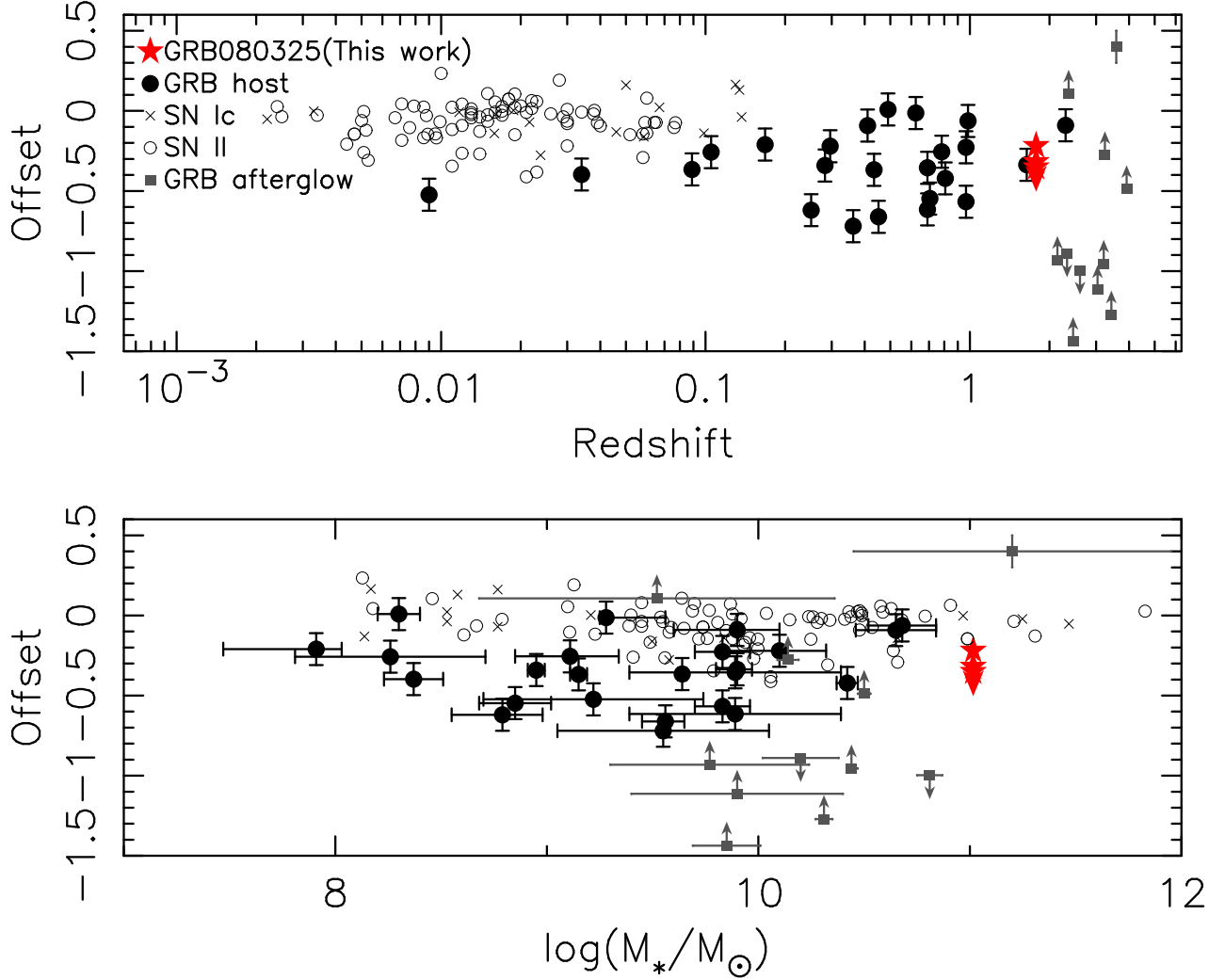}
\caption{
Offset metallicity from normalized-mass metallicity relation as a function of redshift (top) and stellar mass (bottom). 
Symbols are same as figure \ref{mass-metal}.
Note that the offsets for GRB 121024A host and absorption-line metallicities at z $<$ 4 are estimated by comparison with the mass-metallicity relation of star-forming galaxies at z=2.3-3.5 \citep{2008A&A...488..463M,2014ApJ...795..165S}.
Among absorption metallicities collected by \cite{2014arXiv1408.3578C} sample, only host galaxies with known stellar mass are plotted.
}
\label{z-offset}
\end{center}
\end{figure}

\clearpage
\begin{deluxetable}{ccccc}
\tablewidth{0pt}
\tablecaption{Updated ground-based optical photometry of GRB 080325 host galaxy \label{mag}}
\tablehead{
\colhead{Filter} &
\colhead{Mag.} &
\colhead{Error} &
\colhead{System} &
\colhead{Telescope} 
}
\startdata
  $u'$&25.91&0.25 &SDSS&Keck/LRIS \\
    B &25.75&0.12 &Vega&Subaru/Suprime \\
  $g'$&25.48&0.08 &SDSS&Keck/LRIS \\
    R &24.60&0.11 &Vega&Keck/LRIS \\
    R &24.63&0.14 &Vega&Subaru/Suprime \\
  $i'$&24.72&0.15 &SDSS&Keck/LRIS \\
  $i'$&24.67&0.20 &SDSS&Subaru/Suprime \\
  $z'$&24.18&0.09 &SDSS&Subaru/Suprime \\
\enddata
\tablecomments{
Magnitudes are not corrected for Galactic extinction.
}
\end{deluxetable}

\clearpage
\begin{table}
\begin{center}
\caption{Results of SED fitting and emission-line fitting analysis of GRB 080325 host galaxy\label{t_flux}}

\begin{tabular}{ccc}
\tableline\tableline
Stellar mass&SFR$_{\rm SED}$\tablenotemark{a}&$A_{V,}$$_{\rm stellar}$\\
log(M$_{*}$/M$_{\odot}$)&M$_{\odot}$ yr$^{-1}$& \\
\tableline
11.02$^{+0.05}_{-0.09}$&20.3$^{+13.1}_{-9.5}$ &1.17$^{+0.14}_{-0.17}$\\
\end{tabular}

\begin{tabular}{ccccc}
\tableline\tableline
          &H$\alpha$ flux &[NII]$\lambda6584$ flux &Line FWHM&Velocity offset\\
	  &10$^{-17}$ erg s$^{-1}$ cm$^{-2}$ &10$^{-17}$ erg s$^{-1}$ cm$^{-2}$ &km/s&km/s\\
\tableline
Whole     &10.6$\pm$0.34  &2.3$\pm$0.31            &508.4$\pm$27.3       &91.3\\
South part&8.0$\pm$0.22	  &1.3$\pm$0.19		   &465.8$\pm$22.4       &0.0\\
North part&3.0$\pm$0.24	  &$<$0.39\tablenotemark{c}   &565.1$\pm$65.5 	 &474.5\\
South+North\tablenotemark{b}&11.0$\pm$0.34 &2.3$\pm$0.31            &508.0$\pm$21.0       &-\\
\tableline\tableline
          &12+log(O/H)$_{\rm KK04}$\tablenotemark{d} &SFR$_{\rm H \alpha}$\tablenotemark{e} &SFR$_{\rm H \alpha}$\tablenotemark{f}&sSFR \\
          &KK04 calibration         &M$_{\odot}$ yr$^{-1}$                        &M$_{\odot}$ yr$^{-1}$                       &Gyr$^{-1}$\\
\tableline
Whole     &8.88$^{+0.05}_{-0.06}$   &35.6$\pm$1.1                          &47.0$\pm$1.5                         &0.34-0.45\\
South part&8.78$^{+0.05}_{-0.06}$   &27.0$\pm$0.7                          &35.7$\pm$1.0                         &\\
North part&$<$8.75                  &10.0$\pm$0.8                          &13.2$\pm$1.0                         &\\
South+North\tablenotemark{b}&8.70$^{+0.06}_{-0.06}$  &37.0$\pm$1.0         &49.0$\pm$1.3                         &0.36-0.47\\
\tableline\tableline
\end{tabular}

\end{center}
\tablenotetext{a}{SFR derived from SED fitting of GRB 080325 host galaxy.}
\tablenotetext{b}{Sum of South and North part spectra, but North spectra is blue shifted by 474.5km/s.}
\tablenotetext{c}{3-$\sigma$ upper limit.}
\tablenotetext{d}{Metallicity based on PP04 N2 method \citep{2004MNRAS.348L..59P} is converted to KK04 method \citep{2004ApJ...617..240K} using the conversion formula of \cite{2008ApJ...681.1183K} for comparison with other GRB host samples. 
Denoted errors do not include the systematic error ($\sim$ 0.1 dex) associated with metallicity calibration.}
\tablenotetext{e}{$A_{V,}$$_{\rm emission}$ identical to $A_{V,}$$_{\rm stellar}$ derived from SED fitting analysis is assumed. The slit loss of $\sim$ 0.3 is taken into account.}
\tablenotetext{f}{$A_{V,}$$_{\rm emission}$ = $A_{V,}$$_{\rm stellar}$/0.76 \citep{2014ApJ...792...75Z} is assumed. The slit loss of $\sim$ 0.3 is taken into account.}

\end{table}

\end{document}